\documentclass[floats,twocolumn,prb]{revtex4}
\usepackage{amsmath}
\newcommand{\rr}{{\bf r}}
\newcommand{\qq}{{\bf q}}
\newcommand{\Q}{{\bf Q}}
\newcommand{\RR}{{\bf R}}
\newcommand{\kk}{{\bf k}}

\newcommand{\YBCO}{{YBa$_2$Cu$_3$O$_{6+x}$} }

\newcommand{\sigmab}{\overline{\sigma}}
\usepackage{graphicx}
\begin{document}
\title{Superfluid Suppression in $d$-Wave Superconductors due
to Disordered Magnetism} \author{W. A. Atkinson} \affiliation{Trent
University, 1600 West Bank Dr., Peterborough ON, K9J 7B8, Canada}
\date{\today}
\begin{abstract}
The influence of static magnetic correlations on the
temperature-dependent superfluid density $\rho_s(T)$ is calculated for
$d$-wave superconductors.   In self-consistent calculations,
itinerant holes form incommensurate
spin density waves (SDW) which coexist with superconductivity. 
 In the clean limit, the density of states is gapped, and $\rho_s(T\ll T_c)$
is exponentially activated.
In inhomogeneously-doped cases, the SDW are disordered and both the
density of states and $\rho_s(T)$ obtain forms indistinguishable
from those in dirty but pure $d$-wave superconductors, in accordance
with experiments.
We conclude that the observed collapse of $\rho_s$ at $x\approx 0.35$
in underdoped \YBCO may plausibly be attributed to the coexistence of 
SDW and superconductivity.
\end{abstract}
\maketitle

\section{Introduction}
High temperature superconductors (HTS) are an ideal class of materials
with which to study electronic correlations in superconductivity
because the correlations can be tuned from weak to strong via chemical
doping.  A consequence of strong-correlation physics is that the BCS
theory of conventional superconductors fails to describe HTS.  Uemura
demonstrated that, unlike in BCS theory where the superfluid density
$\rho_s$ and critical temperature $T_c$ are independent, HTS exhibit
an approximate scaling $\rho_s \propto T_c$\cite{Uemura}.  The
physical origin of the Uemura relation is not known conclusively but
is consistent with strong-correlation
models\cite{LeeWen97,paramekanti,vanillaRVB} in which the
quasiparticle spectral weight, and consequently $\rho_s$, are
proportional to the hole concentration $p$, where $p$ is measured
relative to the Mott insulator phase.  However, these models fail to
explain both the collapse of superconductivity at a nonzero doping
$p_c \approx 0.05$ and the breakdown of the Uemura relation near
$p_c$, shown by recent experiments in \YBCO (YBCO) \cite{Zuev,broun}.

A number of authors\cite{Emery,Stroud} have suggested that $T_c$ is
governed by phase fluctuations, possibly in combination with
quasiparticle excitations\cite{herbut}, and in particular that the
rapid collapse of $\rho_s$ and $T_c$ at $p_c$ can be thus
explained\cite{iyengar}.  In this work, we examine a completely
different mechanism: the suppression of superfluidity by the formation
of static magnetic moments.  This is motivated by a substantial body
of experimental evidence for the presence of quasistatic magnetism in
underdoped
HTS.\cite{niedermayer,sidis,mook,sanna,lake,panagopoulos,hodges,stock,miller}
There have been previous suggestions that some form of competing order
is important in the underdoped
HTS,\cite{varma,chakravarty,zhang,alvarez} and in particular there
have been numerous studies of the competition between $d$-wave
superconductivity and commensurate
antiferromagnetism.\cite{inui,murakami,kyung,yamase,metzner} These are
generally difficult to reconcile with superfluid density measurements
largely because the competing order introduces an identifiable energy
scale.  Calculations show that this energy scale appears in the
temperature dependence of the superfluid
density\cite{Wu,Micnas,stajic} but such an energy scale has not been
observed experimentally\cite{stajic}.  Indeed, recent microwave
conductivity measurements\cite{broun} of the superfluid depletion,
$\delta \rho_s(T) \equiv \rho_s(0) - \rho_s(T)$, in high quality
single crystals of YBCO find $\delta\rho_s(T) \propto T$ with a
crossover to $\delta \rho_s(T) \propto T^2$ when $T\ll T_c$.  The
linear $T$-dependence
is expected in a
single-phase $d$-wave superconductor and the low-$T$ crossover to $T^2$
behavior has been attributed
to residual impurity scattering.
It is therefore not {\em a priori} clear that the experimentally
observed magnetic moments have any significant effect on the
electronic spectrum.  Here, we show that a phase of coexisting spin
density wave (SDW) and $d$-wave superconducting (dSC) order can,
provided the SDW is disordered, have a spectrum indistinguishable from
that of a dirty dSC.  We conclude that the rapid collapse of
superfluid density near $p_c$ could indeed be due to magnetism.

Our approach is semi-phenomenological.  We construct a mean-field
model in which the model parameters are assumed to have been dressed
by electron interactions. The approach is motivated by a variety of
calculations,\cite{kotliarliu,yoshikazu,ZGRS,paramekanti,LeeWen97,inui}
mostly for the $t$-$J$ model, in which mean-field theories are
developed for which the parameters are functions of $p$.  In the
simplest Gutzwiller approximation for the $t$-$J$ model,\cite{ZGRS}
for example, the renormalized kinetic energy operator $\hat T$ is
related to the bare kinetic energy operator $\hat T_0$ by $\hat T =
\hat T_0 2p/(1+p).$ Other results are found in other
approximations,\cite{kotliarliu,yoshikazu,paramekanti,LeeWen97,inui} but
all show the same qualitative result that $\hat T$ is reduced as one
approaches the Mott insulating phase.  In dynamical mean-field theory
calculations, this derives from a self-energy which renormalizes both
the quasiparticle spectral weight and effective
mass.\cite{kotliarreview}
We remark that the effective interactions are also expected to depend
on the doping, generally increasing as $p$ is reduced.  This is
ignored in our calculations since it will have a quantitative but not
qualitative effect on the outcome.  The essential physics in the
current discussion is that, near the magnetic phase boundary a small change
in the mean-field parameters produces a much larger change in the
magnetic state.  The progression from pure dSC to pure magnetic order
depends only on the general trend that the ratio of kinetic to
interaction energies decreases as $p$ decreases.

In section \ref{model} we introduce the model and describe the phase
diagram.  The most significant result of this section is that it is
possible to have substantial {\em incommensurate} magnetic moments
coexist with the superconductivity, with very little suppression of
the pair amplitude.  This is in contrast with the more widely studied case
of {\em commensurate} magnetic order, which suppresses
superconductivity rapidly.\cite{inui,murakami,kyung,yamase,metzner}  
In section \ref{dos}, we calculate both
the density of states and the superfluid density for the
incommensurate phases.  As mentioned previously, we find the
surprising result that when the magnetic moments are disordered, the
spectrum is indistinguishable from that of a dirty $d$-wave
superconductor.  We conclude briefly in section \ref{conclusion}.

\section{Model and Phase Diagram}
\label{model}
The HTS consist of conducting two-dimensional CuO$_2$ layers that are
weakly coupled along the perpendicular direction.  We model the lower
Hubbard band of a single two-dimensional layer with an extended
Hubbard model, treated at a mean-field level.  Our numerical
calculations have found that the results are only weakly dependent on
the filling $p$ but depend sensitively on the quasiparticle bandwidth
and Fermi surface curvature.  As discussed above, we assume that
doping effects occur indirectly through a parameter $w(p)$ which
renormalizes the quasiparticle dispersion.  For simplicity, and since
the detailed relationship between $w$ and $p$ is not established, we will
treat $w$ as the independent parameter, and keep all other parameters
fixed.

\begin{figure}[tb]
\begin{center}
\includegraphics[width=\columnwidth]{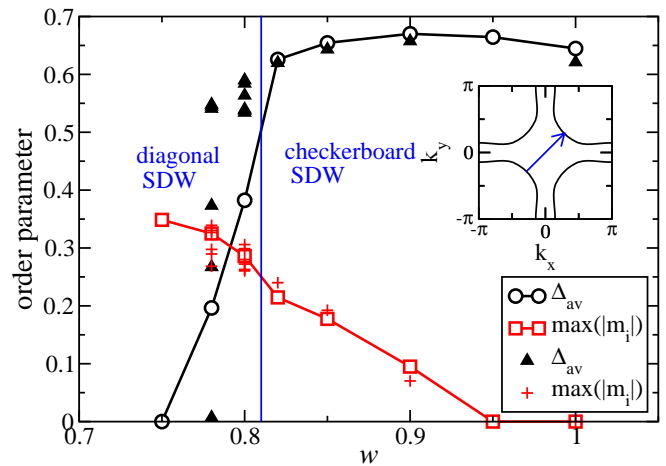}
\end{center}
\caption{(Color online) Phase diagram as a function of kinetic energy
renormalization. Open symbols are for the homogeneously doped
calculations, closed triangle and + symbol are for inhomogeneously
doped calculations.  Inset: Fermi surface and CSDW nesting vector.}
\label{fig1}
\end{figure}

Calculations are for an $N$-site two-dimensional tight-binding lattice
with periodic boundary conditions and lattice constant $a_0=1$,
similar to one used previously to study the local density of states in
underdoped HTS\cite{atkinson}.  The Hamiltonian is
\begin{equation}
\hat H_H = w \sum_{ij\sigma}  t_{ij}c^\dagger_{i\sigma}c_{j\sigma}
+ U \sum_{i\sigma} \hat n_{i\sigma} n_{i\sigmab} + 
\sum_{\langle i,j\rangle} \Delta_{ij} (\hat f_{ij} + \hat f_{ij}^\dagger) 
\end{equation}
where $t_{ij}$ are the hopping matrix elements of the tight-binding
band, $i$ and $j$ are site-indices, $\sigma$ is the electron spin, and
$\sigmab \equiv -\sigma$.  We take $t_{ij}= t_0$ for $i=j$ and
$t_{ij}=t_n$, $n=1,\ldots,3$, for $n$th nearest-neighbor sites $i$ and
$j$.  Taking $\{ t_0,\ldots,t_3\} = \{ 1.7,-1,0.45,-0.1\}$ gives the
Fermi surface shown in Fig.~\ref{fig1}.  The local electronic density
$n_{i\sigma} \equiv \langle \hat n_{i\sigma}\rangle$, where $\hat
n_{i\sigma} = c^\dagger_{i\sigma} c_{i\sigma}$, is determined
self-consistently and the hole density is $p_{i\sigma}=1-n_{i\sigma}$.
The magnetic order parameter is then $m_i =
(n_{i\uparrow}-n_{i\downarrow})/2$, and the staggered moment is $m^Q_i
= (-1)^{x_i+y_i} m_i$ where $\rr_i=(x_i,y_i)$ is the coordinate of
site $i$.  The nearest neighbor pairing term $\Delta_{ij} =
-\frac{J}{2}\langle \hat f_{ij}\rangle$, with $\hat f_{ij} =
(c_{j\downarrow} c_{i\uparrow} - c_{j\uparrow} c_{i\downarrow} )/2$ is
also determined self-consistently and has pure dSC symmetry,
$\Delta_{ij} = \frac{\Delta}{4} (-1)^{y_j-y_i}$, in the nonmangetic
phase.

To allow for inhomogeneous doping we add dopant-impurity and 
Coulomb interaction terms, also treated at the mean-field level:
\begin{equation}
\hat H_c = \sum_{i\neq j} V(\rr_i-\rr_j)\hat n_i n_j - Z\sum_{i} \sum_{\ell=1}^{N_I}
V(\rr_i - \RR_\ell) \hat n_{i},
\end{equation}
with $n_i = \sum_\sigma n_{i\sigma}$, $Z$ the impurity charge,
$\RR_\ell$ the locations of the $N_I$ impurities, and $V(\rr) =
e^2e^{-r/\Lambda}/\epsilon a_0$ a weakly-screened Coulomb interaction,
with $e^2/\epsilon a_0 = 1.5$ and $\Lambda = 20a_0$.  In the cuprate
HTS, donor impurities typically sit a few \AA{} above the CuO$_2$
layers, so we randomly choose values of $\RR_\ell = (x_\ell,y_\ell,d)$
with $d=a_0$.  The total hole doping is $p=ZN_I/N$.  We study a
homogeneously doped case with $N_I=N$ and $Z=p$, and an
inhomogeneously doped case at the same filling with $N_I=N/4$ and $Z =
4p$.  The resulting impurity potential is smoother than one expects in
many underdoped HTS but is reasonable for underdoped YBCO where
approximately 35\% of chain oxygen sites are filled.  It is assumed
that strong-correlation renormalizations of $V$, $J$ and $U$ are
included implicitly and remain constant over the narrow doping range
explored here; for a given $p$, we choose $J$ and $U$ such that $w=1$
corresponds to a pure dSC phase close to the magnetic phase boundary.
We then follow the magnetic phase diagram, Fig.~\ref{fig1}, as $w$ is
reduced.  The results depend sensitively on the Fermi surface shape
(ie.\ on $t_2$ and $t_3$), but depend only weakly on $p$ which is
therefore chosen for computational convenience.  We have studied the
parameter sets $(p,U,J)=(0.05,3.4,1.8)$ and $(p,U,J)=(0.35,3.2,1.5)$.
The two agree semiquantitatively where we have been able to compare;
however, it is difficult to obtain converged solutions for $p=0.05$
when $w$ is small, and we have chosen to present results for $p=0.35$
where a full set of results is available.

\begin{figure}[tb]
\begin{center}
\includegraphics[width=\columnwidth]{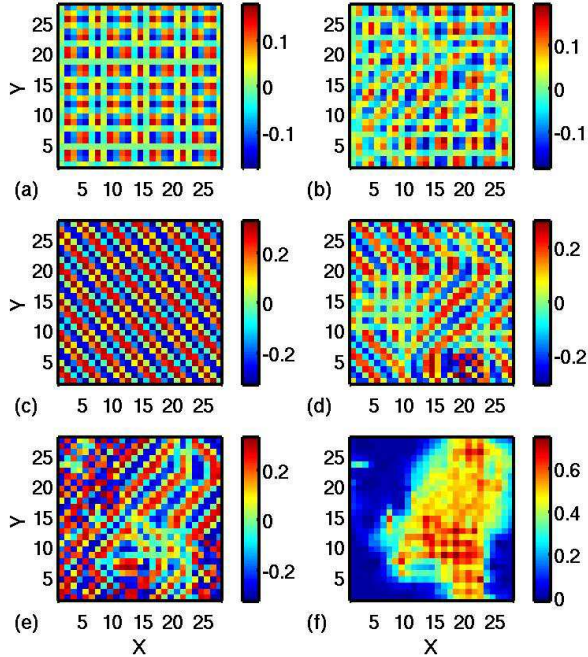}
\end{center}
\caption{(Color online) Typical self-consistent solutions.
The staggered magnetization is shown for 
the CSDW phase at $w=0.85$ (a,b) and the DSDW phase at $w=0.8$
(c,d) with homogeneous (a,c) and
and inhomogeneous (b,d) doping; 
The magnetization (e) and dSC gap (f) are shown for $w=0.78$ with
inhomogeneous doping.  Inhomogeneous results are for a
single dopant configuration.}
\label{fig2}
\end{figure}

The calculations proceed as follows.  The Hamiltonian, written $\hat H
= {\bf c}^\dagger {\bf H} {\bf c}$ with ${\bf c}^\dagger =
[c_{1\uparrow}^\dagger \ldots c_{N\uparrow}^\dagger c_{1\downarrow}
\ldots c_{N\downarrow}]$ and ${\bf H}$ a Hermitian matrix, is
diagonalized numerically giving the unitary matrix ${\bf U}$ of
eigenvectors and eigenvalues ${E_n}$.  The calculations 
\begin{eqnarray}
n_{i\uparrow} &=& \sum_{n,E_n<0} |U_{in}|^2\\ n_{i\downarrow} &=& 1 -
\sum_{n,E_n<0} |U_{i+N n}|^2 \\ \Delta_{ij} &=& -(J/2)\sum_{n,E_n<0}
(U_{in}U_{j+Nn} +U_{i+Nn}U_{jn})
\end{eqnarray}
 are iterated until the largest
difference between successive values of $\Delta_{ij}$ and
$n_{i\sigma}$ is less than $10^{-4}$.  At convergence, the total
energy is typically varying in the tenth significant figure.
Convergence is difficult to obtain: first, the charge density
oscillates wildly in simple-iteration schemes because of the Coulomb
interaction and, second, the magnetic moment configuration may fail to
converge because the energy near self-consistency depends only weakly
on it.  A significant effort has been made to address both these
issues.  First, we have generated our initial guess for the local
charge density using a self-consistent Thomas-Fermi calculation.  In
order not to bias the outcome of the calculation we have used 8-16
randomly seeded initial magnetic moment configurations for each
parameter set, from which the lowest energy self-consistent solution
is retained.  Second, we have adopted a Thomas-Fermi-Pulay iteration
scheme,\cite{pulay} which controls the iteration instability in most
cases.  Results shown here have all converged.

The phase diagram 
is shown in Fig.~\ref{fig1}.  For large $w$, there is a pure dSC
phase.  As $w$ is reduced, there is a second order transition into a
coexisting phase of dSC and checkerboard SDW (CSDW) order at $w
\approx 0.95$, followed by a first order transition into a phase of
coexisting dSC and diagonal SDW (DSDW) order at $w\approx 0.81$; both
phases are illustrated in Fig.~\ref{fig2}.  Superconductivity is
destroyed at $w \approx 0.76$.  From Fig.~\ref{fig2}, one sees that
doping-induced inhomogeneity disorders the SDW and has a significant
effect on the phase diagram in the dSC+DSDW phase: a typical solution
for $w=0.78$, shown in Fig.~\ref{fig2} (e,f), consists of an
inhomogeneous mixture of pure SDW and dSC+SDW order, with
superconductivity preferentially forming in hole-rich regions.  The
spatially-averaged $\Delta$ varies considerably between dopant
configurations, as seen in Fig.~\ref{fig1}, but within superconducting
domains, the local order parameter is consistently $|\Delta_{ij}|\sim
0.5$; the destruction of superconductivity occurs inhomogeneously.

We can understand the origin of magnetic order in the CSDW phase; the
Fourier transform $m(\qq)$ of $m_i$ has a set of four peaks at
$(\pi\pm\delta,\pi\pm\delta)$ and the inset to Fig.~\ref{fig1} shows
that the vector $(\pi-\delta,\pi-\delta)$ taken from the data in
Fig.~\ref{fig2}(a) connects nodal points on the Fermi
surface\cite{nodal}.  The CSDW, therefore, nests portions of the Fermi
surface where the pairing energy is small, and consequently minimizes
competition between magnetic and dSC order.  This explains why
$\Delta$ is roughly constant throughout the dSC+CSDW phase
(c.f. Fig.~\ref{fig1}), and why the transition to the dSC+DSDW phase
appears to occur only when the magnetic energy scale $M \approx U\max(
|m_i|)$ is greater than $\Delta$.  It appears, then, as if CSDW
ordering is stabilized by superconductivity. An analysis of $m(\qq)$
for the DSDW, in contrast, does not reveal any nesting of
high-symmetry points, and $\Delta$ collapses rapidly in the dSC+DSDW
phase.

\begin{figure}
\begin{center}
\includegraphics[width=\columnwidth]{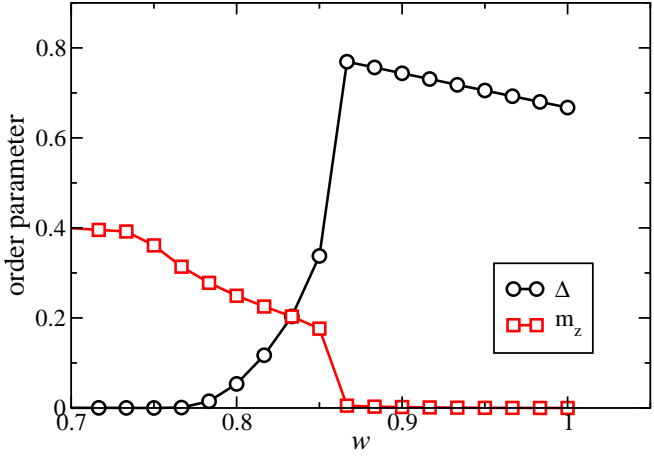}
\caption{(Color online) Phase diagram for commensurate order.  
Here, $U=3.6$ while other parameters are as given in the text.}
\label{fig3}
\end{center}
\end{figure}

These results are in striking contrast to what one finds for the case of
commensurate magnetic order. We show, in Fig.~\ref{fig3} the results
of calculations for the antiferromagnetic moment $m_z$ and the dSC
order parameter $\Delta$ determined self-consistently in the
homogeneous limit.  The calculation proceeds as follows: 
Adopting a four-component notation\cite{kyung}
${\bf A}_\kk \equiv (c_{\kk\uparrow},\, c^\dagger_{-\kk\downarrow},\,
c_{\kk+\Q\uparrow},\, c^\dagger_{-\kk-\Q\downarrow})^T$ where
$c_{\kk\sigma} = N_k^{-1/2}\sum_i c_{i\sigma}\exp(-i\kk\cdot\rr_i)$,
one can write $H=\sum_\kk^\prime {\bf A_k}^\dagger {\bf H_k} {\bf
A_k}$ where the prime indicates a sum over $k_x > 0$ and
\begin{equation}
{\bf H_k} = \left [ \begin{array}{cccc}
\epsilon_\kk & \Delta_\kk & -M & 0 \\
\Delta_\kk & -\epsilon_\kk & 0 & -M \\
-M & 0 & \epsilon_{\kk+\Q} & \Delta_{\kk+\Q} \\
0 & -M & \Delta_{\kk+\Q} & -\epsilon_{\kk+\Q}
\end{array} \right ],
\end{equation}
with the band energy $\epsilon_\kk= t_0 + 2t_1(\cos k_x + \cos k_y) +4
t_2 \cos k_x \cos k_y $ and $\Delta_\kk = \frac{\Delta}{2} (\cos k_x -
\cos k_y )$ and $M = Um_z$.  If the $4\times 4$ matrix diagonalizing
${\bf H}_\kk$ is denoted ${\bf U}_\kk$, then the self-consistent
equations for $\Delta$ and $m_z$ are
\begin{eqnarray}
\Delta &=&-\frac{J}{N_k} \sum_{j=1}^4 \,\sum_{\stackrel{\kk,}{E_{j\kk}<0}}
(\cos k_x - \cos k_y) U_{1j\kk}  U_{2j\kk}, \\
m_z &=& \frac{1}{2N_k} \sum_{j=1}^4  \,\sum_{\stackrel{\kk,}{E_{j\kk}<0}}
[U_{1j\kk} U_{3j\kk} + U_{2j\kk} U_{4j\kk}].
\end{eqnarray}
 Figure \ref{fig3} shows that the dSC and antiferromagnetic order
parameters are generally incompatible, with only a small coexistence
region.  More extensive studies of the phase diagram with commensurate
order by Kyung\cite{kyung} show that the size of the coexistence
region depends on the model parameters, but that the antiferromagnetic
and dSC order always suppress one another.  By contrast, the CSDW
order has very little effect on the dSC phase.

\section{Density of States and Superfluid Density}
\label{dos}

\begin{figure}[tb]
\begin{center}
\includegraphics[width=\columnwidth]{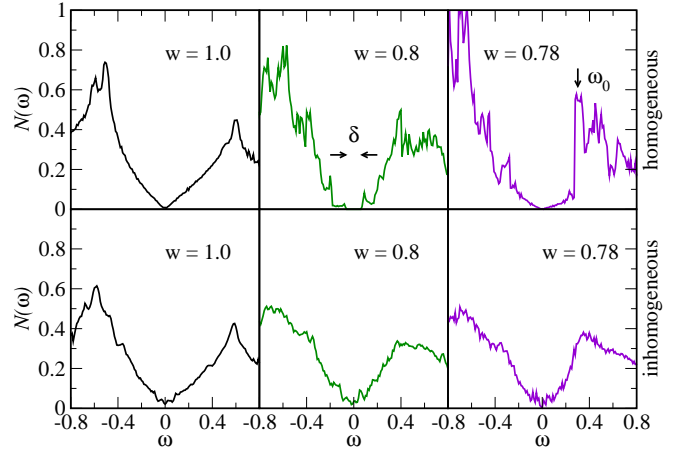}
\end{center}
\caption{(Color online)
Density of states.  $N(\omega)$ is shown for homogeneous (top
row) and inhomogeneous (bottom row) doping.  Inhomogeneous results are
for a single dopant configuration, except for $w=0.78$ which is
averaged over 5 configurations.  The single-particle gap $\delta$ and
subgap resonance at $\omega_0$ are indicated.}
\label{fig4}
\end{figure}

Because of the nodal nesting, the density of states (DOS) 
\begin{equation}
N(\omega) =
\sum_{i=1}^N \sum_{n=1}^{2n} [|U_{in}|^2 \delta(\omega-E_n) +
|U_{i+Nn}|^2\delta(\omega+E_n)],
\end{equation}
 develops a gap of width $\delta$ in
the dSC+CSDW phase, as shown in Fig.~\ref{fig4}.  In contrast, the
dSC+DSDW phase retains the characteristic $d$-wave DOS, $N(\omega)
\propto |\omega|$ at low $\omega$, but develops a resonance at
$\omega_0$ which in many cases dominates the spectrum. In both cases,
these qualitative differences from the pure dSC DOS are reflected in
$\rho_s(T)$.

The superfluid density is related to the magnetic penetration depth
$\lambda(T)$ measured in experiments by $\rho_s(T) = mc^2/4\pi e^2
\lambda^2(T)$.  In linear-response theory, $\lambda^{-2}(T) = (4\pi
e^2/c^2)\langle
K^\mathrm{dia}_{\alpha\alpha}(T)-K^\mathrm{param}_{\alpha\alpha}(T)
\rangle_{\alpha=x,y}$ with
\begin{eqnarray}
K^\mathrm{dia}_{\alpha\beta} &=& 
 \sum_{m} \big [ \tilde M^{-1}_{\alpha\beta} \big ]_{mm} f(E_m) 
\label{dia}\\
K^\mathrm{para}_{\alpha\beta} &=& 
\sum_{m,n} \big[ \tilde \gamma_\alpha \big ]_{mn}
\big[ \tilde \gamma_\beta \big ]_{nm} \frac{f(E_m)-f(E_n)}{E_m-E_n}
\label{para} 
\end{eqnarray}
where
\begin{eqnarray*}
{} [{\bf\tilde M}^{-1}]_{mn} &=& \sum_{i,j} \sum_{p=0}^1 (-1)^p
U^\dagger_{m\,i+pN} [{\bf
M}^{-1}]_{ij} U_{j+pN\,n} 
 \\
{} [{\bf \tilde \gamma}]_{mn} &=& \sum_{i,j} \sum_{p=0}^1 
U^\dagger_{m\,i+pN} [{\bf \gamma}^{-1}]_{ij} U_{j+pN\,n} 
\end{eqnarray*}
where ${\bf \tilde M}$ and ${\bf \tilde \gamma}$ are the inverse
effective mass tensor and current vertex respectively, written in the
basis of eigenstates of the Hamiltonian.  On a tight-binding lattice,
$[ M^{-1}_{\alpha\beta} ]_{ij} = - t_{ij}
\vec\alpha\cdot\rr_{ij}\rr_{ij}\cdot\vec\beta$ and
$[\gamma_\alpha]_{ij} =i\vec\alpha\cdot\rr_{ij}t_{ij}$ with
$\vec\alpha$ and $\vec\beta$ the unit vectors $\hat x$ or $\hat y$ and
$\rr_{ij} = \rr_i-\rr_j$.  The calculations are restricted to low $T$
where we can use the $T=0$ values for $\Delta_{ij}$ and $n_{i\sigma}$;
the $T$-dependence of $\rho_s(T)$ is due to thermal pair breaking, as
has been argued in Ref.~[\onlinecite{LeeWen97}].  We discuss this
assumption below.

\begin{figure}[tb]
\begin{center}
\includegraphics[width=\columnwidth]{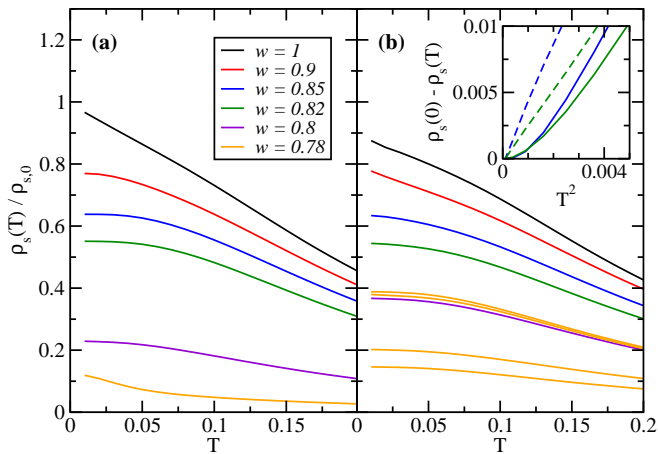}
\end{center}
\caption{(Color online)
Superfluid density. $\rho_s(T)$ is plotted for homogeneous
(a) and inhomogeneous (b) doping.  Different curves for $w=0.78$ in
(b) correspond to different randomly-generated impurity
potentials. Inset: low-$T$ behavior of $\rho_s(T)$ for homogeneous
(solid) and inhomogeneous (dashed) doping.  Results in (b) are
configuration-averaged over 2-4 samples, except for $w=0.78$ where
results are shown for each configuration.}
\label{fig5}
\end{figure}

We focus first on homogeneous doping, Fig.~\ref{fig5}(a).  The
superfluid density at $T=0$ is a strong function of $w$, especially in
the dSC+DSDW phase; $\rho_s(0)$ is reduced to $10\% $ of its initial
value with only small changes in the coherence peak energy in the DOS.
While this is consistent with experiments, the $T$-dependence of
$\rho_s(T)$ is not.  In the dSC+CSDW phase, $\rho_s(T)$ is
exponentially activated: $\delta \rho_s(T) \propto \exp(-T/\delta)$,
where $\delta$ is the single-particle gap shown in Fig.~\ref{fig4}.
In the dSC+DSDW phase, $\rho_s(T)$ is linear in $T$, but also has an
exponential contribution from the resonance at $\omega_0$, also shown
in Fig.~\ref{fig4}.  In most cases we have studied the exponential
contribution is dominant.

The inhomogeneously-doped calculations are qualitatively different.
The electronic potential produced by doping is itself weakly
scattering and has little direct effect on nodal
quasiparticles\cite{Hirschfeld}; however, it disorders the SDW, and
indirectly does have a significant effect on the low-$\omega$ DOS.  As
seen in Fig.~\ref{fig4}, characteristic features of the different SDW
phases are washed out and $N(\omega)$ universally obtains a
dirty-$d$-wave form.  This is one of the main results of this work.
Not surprisingly, $\rho_s(T)$ also obtains the dirty dSC form; as
shown in Fig.~\ref{fig5}(b) and in the figure inset, $\delta \rho_s(T)
\propto T^2$ for $T\ll T_c$, and is linear in $T$ at larger $T$.  In
general, $\rho_s(T)$ varies weakly between disorder configurations.
The notable exception is near the superconducting phase boundary (ie.\
$w=0.78$) where $\rho_s(0)$ depends strongly on disorder
configuration, although $\delta \rho_s(T)$ remains quadratic in $T$.
In macroscopic samples, this will be reflected as a sensitivity to
both sample quality and doping.  The remarkable aspect of
Fig.~\ref{fig5}(b) is that, even for a magnetic energy scale $M \sim
2\Delta$ (at which point $\rho_s(0)$ is near zero), $\rho_s(T)$ has
the appearance of a dirty but pure dSC, as seen in experiments.  This
constitutes our main finding.

The mechanism by which the superfluid density is depleted is quite
interesting.  The diamagnetic response, $K^\mathrm{dia}$, is almost
independent of both $w$ and the magnetic moment: however, the
paramagnetic response at $T=0$, $K^\mathrm{para}(0)$, is a strong
function of the magnetic moment.  This is reminiscent of the
response to disorder in $d$-wave superconductors where Cooper
pair breaking by impurity scattering manifests as a nonzero
$K^\mathrm{para}(0)$.  In this case, however, the broken Cooper pairs
are also apparent as a finite residual density of states at the Fermi
level.  A disorder level sufficient to cause a 90\% reduction in
$\rho_s(0)$, as we have found here, would produce a residual density
of states comparable to that of the normal state.  The fact that such
a residual density of states is {\em not} observed in our case
illustrates that the SDW correlations are not simply breaking Cooper
pairs.

Rather, it is the fact that Cooper pairs in the magnetic phase do not
have a well-defined charge-current which is responsible for the suppression
of $\rho_s(0)$.  In the commensurate (four-band) case, the current operator
is
\begin{eqnarray}
\gamma_x(\kk) &=& \left[\begin{array}{cc}  v_x(\kk) \tau_0 & 0 \\
    0 & v_x(\kk+\Q)\tau_0 \end{array} \right ] 
\end{eqnarray}
with $v_x = \partial \epsilon_\kk /\partial k_x$ and $\tau_0$ the
Pauli matrix.  This matrix is not diagonal in the basis of Bogoliubov
quasiparticles (ie.\ $\tilde \gamma_x(\kk)$ is not diagonal), meaning
that the current is not conserved.  Physically, this is because the
Bogoliubov quasiparticles are formed from mixtures of states with
crystal momenta $\kk$ and $\kk+\Q$.  Then, because $\tilde
\gamma_x(\kk)$ has nonzero off-diagonal matrix elements, there is a
nonzero interband contribution in Eq.~(\ref{para}) which reduces the
overall superfluid density.  We stress that this mechanism is distinct
from either pair-breaking or quasiparticle renormalization (ie.\
strong-correlation) mechanisms for reducing the superfluid density.

We finish with a comment on the relationship between $T_c$ and
$\rho_s(0)$.  An estimate of $T_c$ as the temperature at which a
straight line, fitted to the region $T>0.1$ in Fig.~\ref{fig5}(b),
crosses the $T$-axis yields $T_c \approx 0.36$, for all $w$.  This is
surprising, as it indicates that even for strongly inhomogeneous cases
$T_c$ is determined by the maximum, rather than average, $\Delta$.  It
also suggests that two physical processes neglected in our
calculations, phase fluctuations and glassy SDW dynamics, may play an
important role at higher $T$.  In particular, phase fluctuations are
expected to be pronounced at small $w$ where the superconductivity is
spatially inhomogeneous.  Glassy spin dynamics, provided they remain
slow on electronic time scales, behave as quenched disordering of the
SDW, and should not change our results qualitatively.  An interesting
question, outside the scope of this work, is how the SDW dynamics
affect $\rho_s(T)$ at higher doping, where a gap in the spin-wave
spectrum begins to appear. In summary, it seems likely that, as
suggested in Ref.~\cite{herbut}, $T_c$ is ultimately determined by a
combination of phase and quasiparticle excitations.

\section{Conclusions}
\label{conclusion}
In conclusion, we have shown that incommensurate magnetic correlations
which nest the nodal points of the Fermi surface may coexist with
$d$-wave superconductivity with essentially no suppression of the
superconducting order.  Furthermore, the formation of quasistatic
moments is a plausible explanation for the rapid suppression of
superfluid density near $p_c$ in YBCO.  We find that, provided the
spin density waves are disordered, both the single-particle spectrum
and $\rho_s(T)$ are indistinguishable from the dirty $d$-wave case.

\section*{ACKNOWLEDGMENTS}
This work was performed with the support of Research Corporation grant
CC6062, of NSERC of Canada, and of the Canadian Foundation for
Innovation.  We acknowledge helpful conversations with R. J. Gooding,
M. Randeria, N. Trivedi and Y. Song.


\begin{thebibliography}{niedermayernie}

\bibitem{Uemura} Y. J. Uemura {\em et al.}, Phys. Rev. Lett. {\bf 62}
2317 (1989).

\bibitem{LeeWen97} Patrick A. Lee and Xiao-Gang Wen,
Phys. Rev. Lett. {\bf 78} 4111 (1997).

\bibitem{paramekanti} Arun Paramekanti, Mohit Randeria, and Nandini Trivedi,
Phys. Rev. B {\bf 70}, 054504 (2004).

\bibitem{vanillaRVB} P. W. Anderson, P. A. Lee, M. Randeria,
 T. M. Rice, N. Trivedi, and F. C. Zhang, J. Phys. Condens. Matter {\bf
 16}, 755 (2004).

\bibitem{Zuev} Yuri Zuev, Mun Seog Kim, and Thomas R. Lemberger,
Phys. Rev. Lett. {\bf 95}, 137002 (2005).

\bibitem{broun} D. M. Broun, P. J. Turner, W. A. Huttema, S. Ozcan,
B. Morgan, Ruixing Liang, W. N. Hardy, and D. A. Bonn,
cond-mat/0509223.
 
 \bibitem{Emery} V. J. Emery and S. A. Kivelson, Nature {\bf
374}, 434 (1995). 

\bibitem{Stroud} E. Roddick and D. Stroud, Physica C {\bf 235}, 3696 (1995).




\bibitem{herbut} I. F. Herbut and M. J. Case, Phys. Rev. B {\bf 70},
094516 (2004).

\bibitem{iyengar} M. Franz and A. P. Iyengar, Phys. Rev. Lett {\bf 96},
047007 (2006).

\bibitem{niedermayer} Ch. Niedermayer, C. Bernhard, T. Blasius,
A. Golnik, A. Moodenbaugh, and J. I. Budnick, Phys. Rev. Lett.  {\bf
80}, 3843 (1998).

\bibitem{sidis} Y. Sidis, C. Ulrich, P. Bourges, C. Bernhard,
C. Niedermayer, L. P. Regnault, N. H. Andersen, and B. Keimer,
Phys. Rev. Lett. {\bf 86}, 4100 (2001).

\bibitem{mook} H. A. Mook, Pengcheng Dai, S. M. Hayden, A. Hiess,
J. W. Lynn, S.-H. Lee, and F. Do\u gan, Phys. Rev. B {\bf 66} 144513
(2002).

\bibitem{sanna} S. Sanna, G. Allodi, G. Concas, A. D. Hillier, and R. De Renzi,
Phys. Rev. Lett {\bf 93}, 207001 (2004).

\bibitem{lake} B. Lake, H. M. Rønnow, N. B. Christensen, G. Aeppli,
K. Lefmann, D. F. McMorrow, P. Vorderwisch, P. Smeibidl,
N. Mangkorntong, T. Sasagawa, M. Nohara, H. Takagi and T. E. Mason,
Nature {\bf 415}, 299 (2002); B. Lake, K. Lefmann, N. B. Christensen,
G. Aeppli, D. F. McMorrow, H. M. Ronnow, P. Vorderwisch, P. Smeibidl,
N. Mangkorntong, T. Sasagawa, M. Nohara, and H. Takagi, Nature
Materials {\bf 4}, 658 (2005).

\bibitem{panagopoulos} C. Panagopoulos, J. L. Tallon, B. D. Rainford,
J. R. Cooper, C. A. Scott, and T. Xiang, Solid State Commun. {\bf
126}, 47 (2003); C. Panagopoulos and A. P. Petrovic and A. D. Hillier
and J. L. Tallon and C. A. Scott and B. D. Rainford, Phys. Rev. B {\bf
69}, 144510 (2004).

\bibitem{hodges} J. A. Hodges and Y. Sidis and P. Bourges and
I. Mirebeau and M. Hennion and X. Chaud, Phys. Rev. B {\bf 66},
020501(R) (2002).

\bibitem{stock} C. Stock, W. J. L. Buyers, Z. Yamani, C. L. Broholm, J.-H. Chung, Z. Tun, R. Liang, D. Bonn, W. N. Hardy, and R. J. Birgeneau, Phys. Rev. B {\bf 73}, 100504(R) (2006).

\bibitem{miller} R. I. Miller, R. F. Kiefl, J. H. Brewer,
F. D. Callaghan, J. E. Sonier, R. Liang, D. A. Bonn, and W. Hardy,
Phys. Rev. B {\bf 73}, 144509 (2006).

\bibitem{varma} C. M. Varma, Phys. Rev. Lett. {\bf 83}, 3538 (1999).
\bibitem{chakravarty} S. Chakravarty, R. B. Laughlin, D. K. Morr, and C. Nayak,
Phys. Rev. B {\bf 63}, 094503 (2001).
\bibitem{zhang} S. C. Zhang, Science 1089 (1997).
\bibitem{alvarez} G. Alvarez, M. Mayr, A. Moreo, and E. Dagotto, Phys. Rev.
B {\bf 71}, 014514 (2005).

\bibitem{inui} M. Inui, S. Doniach, P. J. Hirschfeld, and
A. E. Ruckenstein, Phys. Rev. B {\bf 37}, 2320 (1988).

\bibitem{murakami} M. Murakami and H. Fukuyama, J. Phys. Soc. Jpn.
{\bf 67}, 2784 (1998).

\bibitem{kyung} Bumsoo Kyung, Phys. Rev. B {\bf 62} 9083 (2000).

\bibitem{yamase} H. Yamase and H. Kohno, Phys. Rev. B {\bf 69}, 104526
(2004).

\bibitem{metzner} J. Reiss, D. Rhohe, and W. Metzner, cond-mat/0611164.



\bibitem{Wu} K.-K Voo and W. C. Wu Physica C {\bf 417}, 103-109 (2005)

\bibitem{Micnas} B. Tobijaszewska and R. Micnas, Phys. Stat. Sol. {\bf 242}, 468 (2005).

\bibitem{stajic} Jelena Stajic, Andrew Iyengar, K. Levin, B. R. Boyce,
and T. R.  Lemberger, Phys. Rev. B {\bf 68}, 024520 (2003).

\bibitem{kotliarliu} Gabriel Kotliar and Jialin Liu, Phys. Rev. B {\bf
38}, 5142(R) (1988).

\bibitem{yoshikazu} Yoshikazu Suzumura, Yasumasa Hasegawa, and
Hidotoshi Fukuyama, J. Phys. Soc. Jap. {\bf 57} 2768 (1988).

\bibitem{ZGRS} F. C. Zhang, C. Gros, T. M. Rice, and H. Shiba, Supercond.
Sci. Technol., {\bf 1}, 36 (1988).

\bibitem{kotliarreview} Antoine Georges, Gabriel Kotliar, Werner Krauth, 
and Marcelo J. Rozenberg, Rev. Mod. Phys. {\bf 68}, 13 (1996).


\bibitem{atkinson} W. A. Atkinson, Phys. Rev. B {\bf 71}, 024516 (2005).  

\bibitem{pulay} D. Raczkowski, A. Canning, and L. W. Wang, 
Phys. Rev. B {\bf 64}, 121101(R) (2001).

\bibitem{nodal} In $k$-space, the dSC order parameter is $\Delta_\kk =
\Delta(\cos k_x - \cos k_y)/2$, the nodal points are the points on the
Fermi surface where $\Delta_\kk = 0$.  The antinodal points lie at the
intersection of the Fermi surface and the zone boundary, where the gap
obtains its maximum along the Fermi surface.

\bibitem{Hirschfeld} T. S. Nunner, B. M. Andersen, A. Melikyan, and
P. J. Hirschfeld, Phys. Rev. Lett. {\bf 95}, 177003 (2005).

\end{thebibliography}
\end{document}